\documentclass[epj]{svjour}
\usepackage{graphics}
\begin{document}
\title{Localization and Dephasing Driven by Magnetic Fluctuations in Low Carrier 
Density Colossal Magnetoresistance Materials}
\titlerunning{Localization and Dephasing Driven by Magnetic Fluctuations in CMR Materials}	
\author{Eugene Kogan\inst{1,2,}\thanks{\emph{e-mail:} kogan@quantum.ph.biu.ac.il }
\and Mark Auslender\inst{3} \and 
Moshe Kaveh\inst{1,2}}                     
%
%
\institute{Jack and Pearl Resnick Institute of Advanced Technology,
Department of Physics, Bar-Ilan University, Ramat-Gan 52900, Israel 
\and Cavendish Laboratory, Madingley Road, Cambridge CB3 OHE, UK
\and Department of Electrical and Computer Engineering,
Ben-Gurion University of the Negev P.O.B. 653, Beer-Sheva 84105, Israel}
\authorrunning{Kogan et al.}
\date{Received: date / Revised version: date}
%
\abstract{
Localization and dephasing of conduction electrons in a low carrier density 
ferromagnet due to 
scattering on magnetic fluctuations is considered.  
We claim the existence  of  the "mobility edge", 
which separates the states with  fast diffusion and the 
states with  slow diffusion; the latter is determined by the dephasing time. 
When the "mobility edge" crosses the Fermi energy a
large and sharp change of conductivity is observed. The theory provides an
explanation for the observed temperature dependence of conductivity in
ferromagnetic semiconductors and manganite pyrochlores.
} 
\maketitle
\section{Introduction}
\label{intro}
Colossal magnetoresistance (CMR) materials attract nowadays considerable
interest, associated mostly with the properties of double-exchange manganite
perovskites \cite{jin}. Class of CMR materials, however, is much wider
and includes, in particular,  magnetic semiconductors
\cite{nagaev} and manganite pyrochlores \cite{ramirez}. All these materials are
characterized by strong interaction between the localized spins and itinerant
charge carriers. In all these materials 
CMR is associated with the sharp increase of the resistivity  
when the temperature $T$
approaches the Curie temperature $T_c$. 
However, taking into account the large variety of the
materials involved and diverse manifestations of the effect, 
it is difficult to expect that any single theory can provide
a universal explanation of the phenomena. 

We  concentrate on 
low carrier density materials (magnetic semiconductors
and manganite pyrochlores), where  the carriers do not 
affect the spin-spin interaction, and magnetic d- or f-ions interact mainly via
ferromagnetic direct exchange (super-exchange).
These materials
being deficient in chalcogen (oxygen) or being properly doped,
have at low temperatures quasi - metallic conductivity. When the temperature
approaches $T_c$ they undergo the metal-insulator transition.

In the previous publications \cite{kog1,aus,kog4} we suggested, 
that such behavior of the conductivity is due to Anderson 
localization of the carriers driven by spin fluctuations of magnetic ions. 
We considered the spin 
fluctuations as {\it static}; hence the scattering of electrons by the   
fluctuations can be treated as elastic, and hence it 
leads to the existence of the mobility edge $E_c$. (This mechanism is 
close to the phonon scattering induced electron localization 
\cite{gogolin,afonin}.) When the temperature 
increases, so does the scattering  
intensity, which leads to the upward motion of the 
mobility edge. The temperature at which the mobility edge crosses the Fermi 
level is identified with the temperature of the metal-insulator transition (MIT). 
This view point on temperature-induced MIT has also been recognized and legitimated 
in several recent publications \cite{varma,mhardag,li}.

In this paper we consider the influence of the dynamics in the spin subsystem on 
the  transition by developing an idea suggested in Ref. \cite{kog2}

\section{Hamiltonian and Approximations}
\label{sec:1}

The Hamiltonian of the system has the form
\begin{eqnarray}
\label{2ham}
H=\sum_{{\bf k}\sigma}E_k
c^{\dagger}_{{\bf k}\sigma}c_{{\bf k}\sigma} 
-\frac{I}{N}\sum_{{\bf kq}\sigma\sigma'}
\vec{S}_q {\bf \sigma}_{\sigma\sigma'}
c^{\dagger}_{{\bf k}\sigma}c_{{\bf k+q}\sigma'}+H_M, 
\end{eqnarray}
where $E_{\bf k}$ is the bare electron spectrum, $c^{\dagger}_{{\bf k},\sigma}$,
$c_{{\bf k},\sigma}$ are the electron creation and annihilation operators, $I$ 
is the
parameter of Hund exchange between the electrons and localized spins, 
${\bf S}_{\bf q}$ is
the Fourier components of the spin density, ${\bf \sigma}$ is the vector of the 
Pauli spin
matrices and $H_M$ is the direct exchange interaction described by 
super-exchange
integral $J(\bf Q)$.

Let us state the relations between the parameters of the problem. 
We consider the case of wide conduction band $W\gg 2IS$,
where $W\sim 1/ma^2$ is the width of the conduction band 
($a$ is the lattice constant and and $m$ is the electron mass), 
$S$ is the spin of magnetic 
ion. This inequality is certainly applicable to
such magnetic semiconductors as EuO and EuS \cite{cho}. For manganite 
pyrochlores we do not have strong 
inequality \cite{seo}, but we believe that the approximation still works in this 
case, at least semi-quantitatively.\footnote{We
emphasize here  the difference between the manganite pyrochlores,
which are n-type low carrier density  intermediate-band materials, and the 
manganite perovskites, which are p-type high carrier density  narrow-band 
materials; the latter thus being definitely outside the scope of the theory 
presented in the paper.} 
Due to low carrier density considered (not in excess of $10^{-19}$ cm$^{-3}$), 
the Fermi energy $E_F$ is at least an order of
magnitude less then $2IS$ (which is larger than $0.5$ eV in the materials considered
\cite{cho,seo}).
We  consider ferromagnetic phase and temperatures such, that the spin splitting
of the conduction band is larger than $E_F$ (estimations show, that
this will be true  up to the temperatures very close
to $T_c$).  
All our assumptions can be thus reduced to inequalities
\begin{eqnarray}
W\gg 2IS \nonumber\\
2I\overline{S^z}\gg E_F\gg T,
\end{eqnarray}
where $\overline{S^z}$, is 
the average spin of magnetic ion. 

In the wide conduction band case the electron-spin exchange can be treated as 
a perturbation, leading to electron scattering.
The conduction electrons being fully spin-polarized and 
spin-flip processes  thus being forbidden,  the scattering (in the Born
approximation) is connected only with the longitudinal spin correlator
$<\delta S^{z}_{\bf q}\delta S^{z}_{\bf -q}>$. It is argued \cite{patashinskii}, 
that for
the wavevector ${\bf q}$ small enough
($qa< {\rm const}(\overline{S^z})^2$) the correlator is dominated by 
contribution of weakly
interacting spin waves with the dispersion law
\begin{equation}
\omega_{\bf Q}=2\overline{S^z}[J(0)-J({\bf Q})]
\end{equation}
and quasi-classical occupation numbers
\begin{equation}
n_{\bf Q}=\frac{T}{\omega_{\bf Q}}
\end{equation}
As a result the static correlator is  
\cite{patashinskii}
\begin{equation}
\label{corr}
<\delta S^z_{\bf q} \delta S^z_{-{\bf q}}>=
\frac{T^2}{8\overline{S^z}^2C^2}\frac{1}{qa},
\end{equation}
where $C$ is the spin stiffness (for nearest-neighbor exchange in a cubic
lattice $C\simeq T_{c}/2S(S+1)$).

For the transport relaxation time we obtain 
\begin{eqnarray}
\label{approx}
\frac{1}{\tau}=\frac{2\pi}{N}I^2
\sum_{\bf q}<\delta S^z_{\bf q} \delta S^z_{-\bf q}>
\frac{\bf k\cdot q}{k^2}\delta(E_{\bf k}-E_{\bf k+q}) \nonumber\\
=\frac{ma^2I^2 T^2}{16\pi\overline{S^z}^2C^2}
\sim\frac{I^2 S(S+1)}{W}\frac{T^2}{T_c^2}\frac{S^2}{\overline{S^z}^2}
\end{eqnarray} 

We see that for temperatures high enough, $\tau E_F<1$. 
Hence we  need some kind of strong scattering theory.
As such we shall use the self-consistent localization theory by Vollhard and
W\"olfle (VW) \cite{vollhardt}, extended in Ref. \cite{yoshioka} to systems 
without time-reversal invariance. But first we should calculate
the crucial parameter in our approach - the 
dephasing time $\tau_{\varphi}$.

\section{Dephasing Time}
\label{sec:2}

The inverse dephasing time can be defined as the mass of the
Cooperon \cite{altshuller,castellani}. 
(An alternative, but essentially equivalent view on dephasing see in
Ref. \cite{imry}.) For the Cooperon $C({\bf R},t)$ we obtain equation
\begin{equation}
\label{coop}
\left\{\partial/\partial t - D{\bf \nabla}^2+
\left[f(0)-f(t)\right]\right\}C({\bf R},t)=0,
\end{equation}
where
\begin{equation}
\label{f}
f(t)=\frac{2\pi}{N}I^2\sum_{\bf q}\Phi_{zz}({\bf q},t)\delta(E_{\bf k}-E_{\bf k+q})
\end{equation}
and $\Phi_{zz}({\bf q},t)$ is the temporal longitudinal spin correlator
($\Phi_{zz}({\bf q},t=0)\equiv <\delta S^z_{\bf q} \delta S^z_{-{\bf q}}>$).

Eq. (\ref{coop}) can be easily understood if we compare diagrams for the Diffuson
and the Cooperon on Fig. 1. 
The Diffuson  does not have any mass because of Ward
identity. In the case of the Cooperon, the Ward identity is broken: interaction line
which dresses single particle propagator is given by static correlator, and
interaction line which connects two different propagators in a ladder is 
given by dynamic
correlator. The difference $[f(0)-f(t)]$ shows how strongly the Ward identity 
is
broken and, as we'll see below, determines the mass of the Cooperon. 
\begin{figure}
\resizebox{0.5\textwidth}{!}{%
  \includegraphics{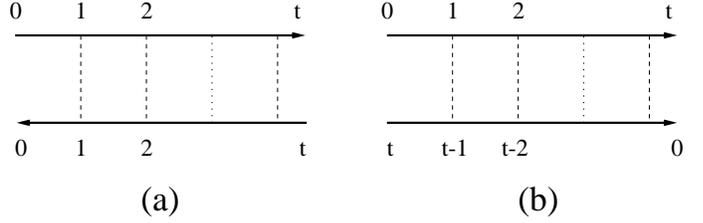}
}
\caption{Diagrams for the Diffuson (a) and the Cooperon (b). Solid line is dressed electron
propagator, dashed line connecting times $t$ and $t'$ corresponds to
$\Phi_{zz}(t-t')$.}
\label{fig:1}       
\end{figure}
Solving Eq. (\ref{coop}) we get \cite{golubentsev}  
\begin{equation}
\label{cooperon}
C(t)=C_{\mathrm{el}}(t)\exp\left\{\int_0^t\left[f(t')-f(0)\right]dt'\right\},
\end{equation}
where $C_{\mathrm{el}}$ is the Cooperon calculated ignoring the in-elasticity 
of scattering.

Using the  spin-wave picture described above, we obtain
\begin{eqnarray}
\label{time}
\Phi_{zz}\left({\bf q},t \right) = 
\frac{1}{N}\sum_{\bf Q}n_{\bf Q}n_{\bf Q+q}
\exp\left[i(\omega _{\bf Q+q}-\omega _{\bf Q})t\right]. 
\end{eqnarray}
Performing integration in Eq. (\ref{cooperon})
we get
\begin{equation}
\label{cooperon2}
C(t)=C_{\mathrm{el}}(t)\exp\left[-t^3/\tau_{\varphi}^3\right],
\end{equation}
where
\begin{equation}
\label{g}
\frac{1}{\tau_{\varphi}^3}=\frac{\pi}{3N^2}I^2\sum_{\bf Q q}
n_{\bf Q}n_{\bf Q+q}\delta(E_{\bf k}-E_{\bf k+q})
(\omega _{\bf Q+q}-\omega _{\bf Q})^2 .
\end{equation}
Calculating the integrals in Eq. (\ref{g}) we obtain
\begin{equation}
\frac{1}{\tau_{\varphi}}=\left (\frac{I^{2}T^{2}({\cal 
W}-1)ma^{5}k_{F}^{3}}{18\pi}\right )^{1/3}\simeq
\left(\frac{I^2T^2E_F^{3/2}}{W^{5/2}}\right)^{1/3}
\end{equation}
where ${\cal W}$ is the Watson integral. It is worth noting that dephasing time is 
defined
by the second time derivative of $\Phi_{zz}({\bf q},t)$ at $t=0$ which can be
calculated via second moment of corresponding spectral density; the result turns 
out to be essentially the same as Eq. (\ref{g}). So the spin-wave picture, being 
physical one, is not crucial for obtaining $1/\tau_{\varphi}$.

It should be noticed that the form of the Eq. (\ref{cooperon2}) for the 
Cooperon is quite general,
provided the scatterers are in a ballistic motion, irrespective of whether they
are point particles \cite{golubentsev}, phonons \cite{afonin}, or
spin waves, like in our case.   

The result  for the dephasing time 
can be understood using simple qualitative arguments. 
If all the collisions  lead to the same 
electron energy change $\delta E$,
the dephasing time could be obtained using relation \cite{altshuller}
\begin{equation}
\tau_{\varphi}\delta E\sqrt{\frac{\tau_{\varphi}}{\tau_{\mathrm{out}}}}\sim 
2\pi, \end{equation}
where 
$\tau_{\varphi}/\tau_{\mathrm{out}}$ is just the number of scattering acts 
during the time  ${\tau_{\varphi}}$ ($\tau_{\mathrm{out}}$ is the 
extinction time). So in this case 
\begin{equation}
\label{toy}
\frac{1}{\tau_{\varphi}^3}\sim\frac{(\delta E)^2}{\tau_{\mathrm{out}}}.
\end{equation}
If we rewrite the formula for the 
extinction time
\begin{eqnarray}
\label{ext}
\frac{1}{\tau_{\mathrm{out}}}=\frac{2\pi}{N}I^2
\sum_{\bf q}<\delta S^z_{\bf q} \delta S^z_{-\bf q}>
\delta(E_{\bf k}-E_{\bf k+q})
\end{eqnarray} 
in the form
\begin{equation}
\label{rewrite}
\frac{1}{\tau_{\mathrm{out}}}=\frac{2\pi}{N^2} I^2\sum_{\bf Q q}n_{\bf Q}n_{\bf 
Q+q}
\delta(E_{\bf k}-E_{\bf k+q})
\end{equation} 
and notice that $(\omega _{\bf Q+q}-\omega _{\bf Q})$ is just the energy
change of the electron when scattering on a spin wave, we immediately see that
Eq. (\ref{g}) is just Eq. (\ref{toy}) with the 
integration with respect to different collision induced energy changes built 
in. 

\section{Conductivity Calculation}
\label{conductivity}

The time-reversal invariance in the system we are considering is broken for two 
reasons. First, 
because we are considering ferromagnetic system, it is naturally to 
expect that   the magnetic field is present in the system. Even more important 
is that  the dephasing itself breaks the time-reversal invariance. We have 
shown 
in the previous Section, that due to dephasing the diffusion pole of the 
particle-particle 
propagator disappears, although particle-hole propagator still has a diffusion 
pole, which is guaranteed by particle number conservation.  
Inserting Eq. (\ref{cooperon2}) into the self-consistent equations proposed in 
Ref. \cite{yoshioka}, for the (particle-hole) diffusion coefficient $D$ and
the particle-particle diffusion coefficient $\tilde{D}$ we obtain system
\begin{eqnarray}
\label{return}
\frac{D_0}{D}=1+\frac{1}{\pi\nu}\int_0^{\infty}\sum_{\bf k}
e^{-\tilde{D}\left({\bf k}+\frac{2e}{c}{\bf A}\right)^2t 
-t^3/\tau_{\varphi}^3}dt 
\end{eqnarray} 
\begin{eqnarray}
\label{ret}
\frac{D_0}{\tilde{D}}=1+\frac{1}{\pi\nu}\int_0^{\infty}
\sum_{\bf k}e^{-Dk^2t} dt,
\end{eqnarray} 
where  $\nu$ is the density of states at the Fermi surface, 
$D_0$ is the diffusion coefficient calculated 
in Born approximation and the momentum cut-off $|{\bf
k}|<1/\ell$, where $\ell$ is the 
transport mean free path,  is implied. 
The conductivity is connected to the diffusion coefficient in a usual way
\begin{equation}
\label{sigma}
\sigma =ne^2(3D/2E),
\end{equation}
where $E$ is the Fermi energy, and $n$ is the concentration. 

For simplicity, we will make an analysis of self-consistent equations only in 
the absence of magnetic field ($A=0$). 
In our case ($\tau_{\varphi}\gg\tau$), like in the case of purely elastic 
scattering, the conductivity
drastically differs in the regions $E>E_c$ and $E<E_c$, where the mobility edge
$E_c$ is obtained from the equation \cite{vollhardt}
\begin{equation}
\label{mit}
E_c\tau=\sqrt{3/4\pi}.
\end{equation} 
More exactly, we have essentially three regions:

\noindent
1. metallic region ($E>E_c$)  with fast diffusion 
\begin{equation}
D\sim D_0,
\end{equation}
where dephasing is irrelevant;

\noindent
2. "dielectric region" ($E<E_c$)  with slow diffusion
\begin{equation}
\label{diel}
D\sim D_0(k\ell)^2(\tau/\tau_{\varphi}),
\end{equation}
determined by the dephasing time;

\noindent
3. critical region around $E_c$,  
($|E/E_c-1|\ll \left(\tau/\tau_{\varphi}\right)^{1/3}$)
\begin{equation}
\label{diel2}
D\sim D_0(\tau/\tau_{\varphi})^{1/3}.
\end{equation}

When  the "mobility edge" crosses the Fermi level (it is achieved by tuning the 
temperature) the resistivity changes sharply, 
which looks like a metal-insulator transition.

If we want to take into account the magnetic field in Eq. (\ref{return}), 
it must be noticed, that the vector potential 
${\bf A}$ does not commute with the momentum ${\bf k}$. So the equation takes 
the form
\begin{equation}
\label{cutt}
\frac{D_0}{D}=1+\frac{1}{2\pi\nu}\int_0^{\infty}\sum_{{\cal E}_{\bot}}
e^{-{\cal E}_{\bot}t-t^3/\tau_{\varphi}^3}dt 
\end{equation}
where 
\begin{equation}
{\cal E}_{\bot}=\tilde{D}\left[k_H^2+\frac{4}{l_H^2}\left(N+\frac{1}{2}\right)\right]
\end{equation}
($l_H$ is the magnetic length). 

\section{Discussion}

Let us return to Eq. (\ref{g}). The 
electron energy change in a single scattering 
$\delta E \sim T_c\sqrt{E_F/W}\ll T$, though all the spin waves (with the 
 energies up to $\sim \overline{S^z}T_c/S$) participate
in the dephasing.  This quasi-elasticity 
of scattering gives the opportunity to calculate the dephasing time the 
way we did. (The quasi-elasticity condition holds even better for  Eq. 
(\ref{rewrite}); in this case only the spin waves with small wave vectors 
contribute.)  

When analyzing explicitly the CMR effect, we should first and most take into 
account the influence of the magnetic field on the spin disorder.
The static spin correlator (in ferromagnetic phase) becomes \cite{patashinskii}
\begin{equation}
\label{corr2}
<\delta S^z_{\bf q} \delta S^z_{\bf -q}>=\frac{T^2}{4\pi\overline{S^z}^2C^2}\frac{1}{qa}
\tan^{-1}\frac{q\xi}{2}, 
\end{equation}
where $\xi\sim a\sqrt{\overline{S^z}C/g\mu _{B}H}$ is the correlation length. 
Thus the long wave 
spin fluctuations are suppressed, which  decreases scattering and hence 
reduce the mobility edge.
This mechanism is appropriate for describing CMR effect 
in magnetic semiconductors \cite{aus}, and can be applied to  
manganite pyrochlores (these results will be presented elsewhere).  

Second, magnetic field  shifts the mobility edge 
by cutting off the Cooperon 
(see Eq. (\ref{cutt})). It is appropriate here to explain, why
dephasing, which also cuts off the
Cooperon, influences the localization in a totally different way.   
Consider a case of no magnetic field and a simplified version of the 
self-consistent localization 
theory, when we 
ignore the difference between $D$ and $\tilde{D}$, and also consider the 
dephasing mechanism which leads to
$C(t)=C_{\mathrm{el}}(t)\exp\left[-t/\tau_{\varphi}\right]
$ time dependence.\footnote{This happens when the scatterers
are in a diffusive motion \cite{golubentsev}.}
Then instead of  equations (\ref{return}) and (\ref{ret}) we have a single one  
\begin{eqnarray}
\label{retu}
\frac{D_0}{D}=1+\frac{1}{\pi\nu}\sum_{\bf k}\frac{1}{Dk^2+1/\tau_{\varphi}}. 
\end{eqnarray} 
We see, that due to presence of 
$k^{d-1}$ in the numerator in this equation, 
the pole of the Cooperon is of no 
special importance at $d=3$. The dephasing leads to the 
delocalization not because it leads to the disappearance of the diffusion pole, 
but because there appears  in the denominator the term, which does not depend on 
$D$.  

Consider finally the paramagnetic (PM) phase. In the absence of self-consistent 
localization theory which takes into account the spin-flip processes, 
a very rough idea about the localization  in the PM phase  
we can get  from the Ioffe-Regel criterium for the position of the mobility edge 
$\tau E_F\approx 1$. Using the well-known expression for spin-disorder scattering rate 
at temperatures above, but not too close to, $T_c$ we arrive to two opportunities. 
For the relatively high Fermi energy
$E_F>E_0\sim I^4S^4/W^3$
the increase of the temperature above $T_c$ leads to a reverse 
insulator-metal transition. In the opposite case the system remains 
in the dielectric phase.

\begin{acknowledgement} 
 We are grateful to D. Khmel'nitskii for the  useful 
discussions.
\end{acknowledgement}

\end{document}